\documentclass[%
 reprint,
superscriptaddress,
preprintnumbers,
 amsmath,amssymb,
 aps,nofootinbib
]{revtex4-2}

\RequirePackage[running]{lineno}

\usepackage{graphicx}  
\usepackage{ulem}
\usepackage{dcolumn}   
\usepackage{bbm}        
\usepackage{amsmath}

\usepackage{amsfonts}
\usepackage{amssymb}   
\usepackage{epstopdf}
\usepackage{slashed}
\usepackage{multirow}
\usepackage{color}
\usepackage{float}

\usepackage{verbatim}

\usepackage[compat=1.1.0]{tikz-feynman}
\usepackage[colorlinks=true,
            linkcolor=blue,
            citecolor=blue,
            urlcolor=blue,breaklinks=true]{hyperref}
\usepackage{cleveref}

\def\gsim{\lower0.5ex\hbox{$\:\buildrel >\over\sim\:$}}
\def\lsim{\lower0.5ex\hbox{$\:\buildrel <\over\sim\:$}}

\definecolor{wildstrawberry}{rgb}{1.0, 0.26, 0.64}

\newcommand{\be}{\begin{equation}}
\newcommand{\ee}{\end{equation}}
\newcommand{\bea}{\begin{eqnarray}}
\newcommand{\eea}{\end{eqnarray}}

\newcommand{\nbox}{{\,\lower0.9pt\vbox{\hrule \hbox{\vrule height 0.2 cm
\hskip 0.2 cm \vrule height 0.2 cm}\hrule}\,}}

\def\ie{{\it i.e.}}

\def\sub#1{_{\lower.25ex\hbox{$\scriptstyle#1$}}}

\def\gev{\,{\rm GeV}}

\def\ev{\,{\rm eV}}
\def\to{\rightarrow}
\def\slash{\not\!}

\newskip\zatskip \zatskip=0pt plus0pt minus0pt
\def\matth{\mathsurround=0pt}
\def\lsim{\mathrel{\mathpalette\atversim<}}
\def\gsim{\mathrel{\mathpalette\atversim>}}
\def\sigv{\ifmmode \langle\sigma v\rangle\else $\langle\sigma v\rangle$\fi}
\newskip\zatskip \zatskip=0pt plus0pt minus0pt
\def\matth{\mathsurround=0pt}
\def\lsim{\mathrel{\mathpalette\atversim<}}
\def\gsim{\mathrel{\mathpalette\atversim>}}
\def\atversim#1#2{\lower0.7ex\vbox{\baselineskip\zatskip\lineskip\zatskip
  \lineskiplimit
  0pt\ialign{$\matth#1\hfil##\hfil$\crcr#2\crcr\sim\crcr}}}

\def\missET {{\slash{p}_\textrm{T}}}

\begin{document}

\thispagestyle{empty}

\vspace{0.5in}

\title{Time-dependent signals of new physics at the LHC}
\author{Max H. Fieg}
\affiliation{Particle Theory Department, Fermilab, Batavia, IL 60510 USA}
\author{Patrick J. Fox}
\affiliation{Particle Theory Department, Fermilab, Batavia, IL 60510 USA}
\author{Jinbo Zhang}
\affiliation{Department of Physics \& Astronomy, University of California, Irvine 92617}
\author{Aishik Ghosh}
\affiliation{School of Physics, Georgia Institute of Technology, Georgia, USA, 30332}
\author{Virat Varada}
\affiliation{University High School, Irvine, CA 92612}
\author{Daniel Whiteson}
\affiliation{Department of Physics \& Astronomy, University of California, Irvine 92617}

\begin{abstract}

The Large Hadron Collider (LHC) is sensitive to signals of beyond the Standard Model physics through a variety of channels including missing energy and resonance searches. 
In most searches, the new physics and the Standard Model backgrounds are assumed to be invariant in time, up to systematic effects from the experiment. However, new physics with a time variation would provide an additional handle to separate signal from background.
Such a time variation may come from ultralight dark matter coupling to an oscillating background field.
In this paper, we consider an interaction of dark matter with quarks and an additional heavy particle, and show that the sensitivity of a search that uses timing information at the LHC can be up to a factor of two stronger compared to one that does not use time information. 

\end{abstract}
\preprint{FERMILAB-PUB-26-0234-T}
\maketitle

\section{Introduction}

The Large Hadron Collider (LHC) has been a crucial instrument to test the validity of the Standard Model (SM). Despite this success, there has, so far, been no convincing signal of new physics with existing search strategies. These searches typically leverage the predicted kinematic distribution to isolate signal from background---for example, a localized resonance in an invariant mass distribution or large missing transverse momenta in the tail of a distribution. While searches have thoroughly explored these kinematic spaces, new physics which is time-dependent has not been extensively considered. A recent study~\cite{Bigaran:2025uzn} explored exotic time-dependent lepton number violating decays at Mu3e, Belle-II, and the proposed FCC-ee. Similarly time-dependent signals at the LHC provide a new potential avenue for discovery, particularly in kinematic regions that have a large, time-independent, background.

Theoretically, signals of new physics that exhibit intrinsic time dependence are highly motivated in the context of LHC physics. The most notable example arises from interactions with a background ultralight dark matter (DM) candidate~\cite{Jaeckel:2022kwg,Hui:2016ltb,Hui:2021tkt,Eberhardt:2025caq}; some examples include the QCD axion or axion-like particle~\cite{Graham:2013gfa,Preskill:1982cy,Dine:1982ah}, dark photon~\cite{Holdom:1985ag,Fabbrichesi:2020wbt,Nelson:2011sf,Arias:2012az}, or a dilaton~\cite{Banerjee:2025uwn}. For these types of models, the DM component is typically well below the eV-scale, 
such that its de Broglie wavelength, $\lambda_{\rm DM} = 2\pi/m_{\rm DM}~v_{\rm DM}\approx 10^9 ~{\rm km}\times (10^{-15} ~{\rm eV}/m_{\rm DM})$, is macroscopic. In this case, the DM can be treated as a coherent background field that oscillates with a period $T = 2\pi / m_{\rm DM}\approx 4{\rm s}\times (10^{-15} ~{\rm eV}/m_{\rm DM})$. If the DM couples to the SM, then it generically imparts time-dependence in a number of experiments that search for temporal modulation of physical observables.

It is possible that the ultralight DM mediates an interaction between the SM and a dark sector that may contain a heavy TeV-scale particle. If the interaction is restricted to this case, then the sensitivity of most other searches for ultralight DM will be suppressed because they take place at relatively low energy. In this scenario, resonance and missing momentum searches for such a TeV-scale particle would exhibit a time-dependent component due to the oscillating background DM field, and the LHC would be a natural place to perform such a search.

When the signal is time-dependent and the background is not, their differing behavior provides additional discriminating information, orthogonal to kinematic information, that can boost the discovery or exclusion power of searches. For example, an oscillating signal creates intervals of higher and lower signal-to-background ratio, which directly enhances statistical power without additional kinematic requirements that would dilute the signal. It is also possible that new particles have a mass that is dependent on the DM~\cite{Bauer:2026dgu}, in which case the resonance signal will be distributed across multiple invariant mass bins, hiding the signal.

In this paper, we explore time-dependent signatures at the LHC. We first discuss an example model that realizes an interaction between the ultralight DM field, a heavy TeV-scale particle, and the SM. We then present the limits that can be achieved on new physics from an existing jet+$\missET$ analysis~\cite{ATLAS:2021kxv} assuming the signal oscillates with time.  We next estimate the statistical enhancement of time-dependent analysis in the context of resonance searches where the temporal behavior is known or assumed. Then, we demonstrate an application of the anomaly-detection technique CATHODE~\cite{Hallin:2021wme}, which leverages data features motivated by the expected time-dependence of the DM signal to learn this time-dependence directly, thereby providing additional discrimination and discovery power. Finally, we demonstrate the application of sidebands in both time and invariant mass, which allows for a direct measurement of the background within the signal region. Then we conclude with a discussion on potential backgrounds and ideas for further work. 

\section{Model}
\label{ref:model}

Ultralight bosonic DM ($m_{\rm DM}\lsim 0.1\ev$) has a large phase space occupancy and its behavior is that of a classical field~\cite{Hui:2021tkt,Antypas:2022asj}, which oscillates with period $\sim 2\pi/m_{\rm DM}$.  This time variation can rear its head at the LHC in many ways, depending on the form of the couplings of the dark matter to the SM.  For instance, fundamental parameters of the SM Lagrangian \cite{Brzeminski:2020uhm,Safronova:2017xyt,Berlin:2016woy,Brdar:2017kbt,Krnjaic:2017zlz,Capozzi:2018bps,Losada:2023zap,Dine:2024bxv,Zhang:2023lem,Alda:2024xxa} or masses of new particles \cite{Guo:2022vxr,Dev:2022bae,Bauer:2026dgu} can acquire a time dependence. We will be interested in the case where the rate of production of new physics varies with time, while the SM processes and the masses of all states are constant in time.  We are motivated by the model analysed in \cite{Bigaran:2025uzn} to investigate time dependence in charged lepton flavor violation, such as $\tau(\mu)\to e + \mathrm{invisible}$, but will investigate a coupling of the dark sector to quarks in a flavor diagonal way.  The coupling between the dark sector and the SM is often taken to be linear in the DM field, but it can be that the leading order coupling is quadratic \cite{Bouley:2022eer,Banerjee:2022sqg,Bigaran:2025uzn,Delaunay:2025pho}, which is the case we consider.

We are interested in the case where the ultralight DM ($\chi$) and a more massive scalar from the dark sector ($\phi$) couple to a quark current, 
\begin{equation}
    \label{eq:lagrangian}
    {\cal L}\supset\frac{1}{\Lambda^2} \chi~\partial_\mu\phi~ \bar{q}\gamma^\mu q~.
\end{equation}
Such a coupling between ultralight DM and the SM typically introduces fine-tuning issues due to loop corrections of the $\chi$ mass.  These issues can be ameliorated through introduction of multiple copies of the SM \cite{Hook:2018jle,Banerjee:2025zcd,Delaunay:2025pho} or by making the DM a pseudo-Nambu-Goldstone boson (pNGB).  In particular, we follow the approach of \cite{Bigaran:2025uzn} which introduced non-Abelian pNGBs (npPNGBs) from a broken $SU(3)_L \times SU(3)_R$ in the dark sector that is broken to $SU(3)_V$.  In addition, the dark quark masses are taken to be hierarchical with $m_{u'}\gg m_{d'},m_{s'}$ making the dark kaons much lighter than all other npPNGBs.  Charging some SM quarks, as well as the dark sector quarks, under a spontaneously broken $U(1)'$ results in the operator in \Cref{eq:lagrangian} after integrating out the dark photon.  Consequently, $\Lambda \sim v'$ where the dark Higgs vacuum expectation value, $v'$, sets the dark photon mass.  

For the LHC phenomenology we will be interested in, the strong coupling scale in the dark sector and the dark up-quark mass will both need to be TeV scale.  More generally, the above interaction could also appear if $\phi$ and $\chi$ are the physical states after diagonalizing a mass matrix of fields that are charged under a broken ${\rm U}(1)$ to result in an off-diagonal interaction, although some fine-tuning would likely be required to maintain the mass hierarchy. Whatever the origin of this interaction, we use the time-modulation that it predicts for missing energy and resonance signals; the kinematic distributions from this operator are only used in \Cref{sec:knownosc} and \Cref{sec:learnedosc}. Other interactions involving a TeV scale particle and an ultralight DM will likely have similar phenomenological implications for studying time-dependent signatures at the LHC.

The DM, $\chi$, is taken to be ultralight such that there is a large occupation number density, and thus takes the form of a classical background field.  For a set of dark matter particles of speed $v$ this classical field behaves as $\sin (m_\chi(1+v^2/2)t + \delta)$.  The local DM speeds are distributed according to (typically a Maxwell Boltzmann) distribution $f(v)$.  Coarse graining this distribution over $N$ bins of width $\Delta v$ up to the escape velocity ($v_{esc}\sim 700$ km/s), the classical DM field in our local vicinity is taken to be~\cite{Foster:2017hbq}
\begin{align}
\label{eq:timedep}
    \chi(t) &= 
    \frac{\sqrt{f_{\rm DM}\rho_{\rm DM}}}{m_\chi}
    \sum_{j=1}^N \alpha_j \sqrt{f(v_j)\Delta v} \nonumber \\
    &\quad\times 
    \sin\!\left[m_\chi \left(1+\frac{v_j^2}{2}\right)t + \delta_j \right]
\end{align}
where $\rho_{\rm DM}\approx0.4~{\rm GeV/cm^3}$~\cite{Catena:2009mf} is the average DM energy density, and $f_{\rm DM}$ is the fraction of the dark matter that is in the ultralight field. The coefficients $\alpha_j$ describe the random amplitudes of individual velocity modes and are drawn from the Rayleigh distribution. Recent studies of dwarf galaxies~\cite{Zimmermann:2024xvd} bound the mass of a dominant DM component to be $m_\chi\gtrsim 2.2\times10^{-21}~{\rm eV}$, resulting in an oscillation period of $T_{\rm Period} = 2\pi/m_\chi \lesssim 21.7~ {\rm days} \times (2.2\times10^{-21} ~{\rm eV}/m_\chi)$. This description results in a characteristic coherence time, $T_{\rm cohere.} \approx T_{\rm Period} /\langle v \rangle^2 = T_{\rm Period}\times 10^6$; for observation times longer than the coherence time, the amplitude modulates and for shorter observation times \Cref{eq:timedep} is well-approximated by a single sine wave.

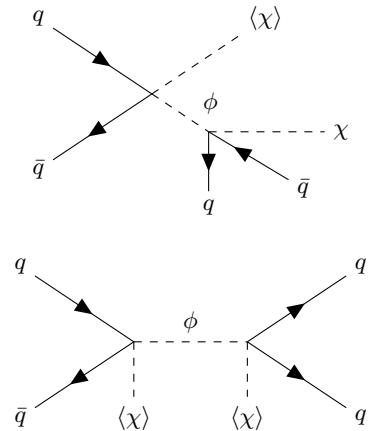
\begin{figure}[t]
\centering

\begin{tikzpicture}
  \begin{feynman}
    \vertex (i1) at (0, 1) {\(q\)};
    \vertex (i2) at (0, -1) {\(\bar{q}\)};
    \vertex (v1) at (1.5,0);
    \vertex (v2) at (3, 1) {\(\langle\chi\rangle\)};
    \vertex (f11) at (2.25, -1.5) {\(q\)};
    \vertex (f12) at (4.0, -0.5) {\(\chi\)};
    \vertex (f13) at (3.5, -1.25) {\(\bar{q}\)};
    \vertex (f1) at (2.25, -0.5);

    \diagram* {
      (i1) -- [fermion] (v1),
      (i2) -- [anti fermion] (v1),
      (v1) -- [scalar] (v2),
      (v1) -- [scalar, edge label={\(\phi\)}, near end] (f1),
      (f1) -- [scalar] (f12),
      (f1) -- [fermion] (f11),
      (f1) -- [anti fermion] (f13)
    };
  \end{feynman}
\end{tikzpicture}

\vspace{1em} 

\begin{tikzpicture}
  \begin{feynman}
    \vertex (i1) at (0, 1) {\(q\)};
    \vertex (i2) at (0, -1) {\(\bar{q}\)};
    \vertex (v1) at (1.5, 0);
    \vertex (v2) at (3, 0);
    \vertex (f1)  at (4.5, -1) {\(q\)};
    \vertex (f2)  at (4.5,  1) {\(q\)};
    \vertex (f3)  at (3, -1) {\(\langle\chi\rangle\)};
    \vertex (f4)  at (1.5, -1) {\(\langle\chi\rangle\)};

    \diagram* {
      (i1) -- [fermion] (v1),
      (i2) -- [anti fermion] (v1),
      (v1) -- [scalar, edge label={\(\phi\)}] (v2),
      (v2) -- [fermion] (f1),
      (v2) -- [fermion] (f2),
      (v2) -- [scalar] (f3),
      (v1) -- [scalar] (f4)
    };
  \end{feynman}
\end{tikzpicture}

\caption{Examples of time-dependent processes of the interaction in \Cref{eq:lagrangian}. Top: missing-energy signature with a $\chi^2(t)$ dependence; Bottom: dijet-resonance signature with a $\chi^4(t)$ dependence. Here, $\langle \chi\rangle$ is the classical expectation value of the background DM field. 
}
\label{fig:diagrams}
\end{figure}

The interaction in \Cref{eq:lagrangian} allows for a number of time-dependent signatures of new physics at proton collisions at the LHC. There is a time-dependent missing energy signature associated with the radiation and subsequent decay of $\phi$ which exhibits a $\chi^2(t)$ dependence, as well as a time-dependent $s$-channel resonant production of $\phi$ with a $\chi^4(t)$ dependence, as shown in \Cref{fig:diagrams}. All of the relevant processes can be made time-independent by taking the expected value of the DM field to be an on-shell production in the final state, \textit{i.e.} $\langle \chi\rangle\rightarrow\chi$. In general, such time-independent processes would be phase-space suppressed due to the additional final-state particles, and the expectation value of $\chi$, $\langle\chi\rangle=\sqrt{2\rho_{\rm DM}}/m_\chi$, can be an enhancement for small masses (\ie ~large occupation number).

For sufficiently large masses (small $T_{\rm Period}$), the coherence length may become comparable to the LHC vertex resolution, $\mathcal{O}(10\,\mu{\rm m})$~\cite{ATLAS:2016nnj}. In that case, different coherence patches would correspond to different $\{\alpha_j,\delta_j\}$, potentially reducing the event rate and washing out the time dependence if the vertex cannot be resolved. In practice, this is not a limitation: a coherence length smaller than the vertex resolution implies $T_{\rm Period}\lesssim  10^{-14}\,\mathrm{s}$, which is
well below the LHC bunch spacing, the detector timing resolution, and the oscillation periods considered in this work.

\section{Time-dependent missing momentum}

\begin{figure*}[t]
  \centering
  \includegraphics[width=0.95\textwidth]{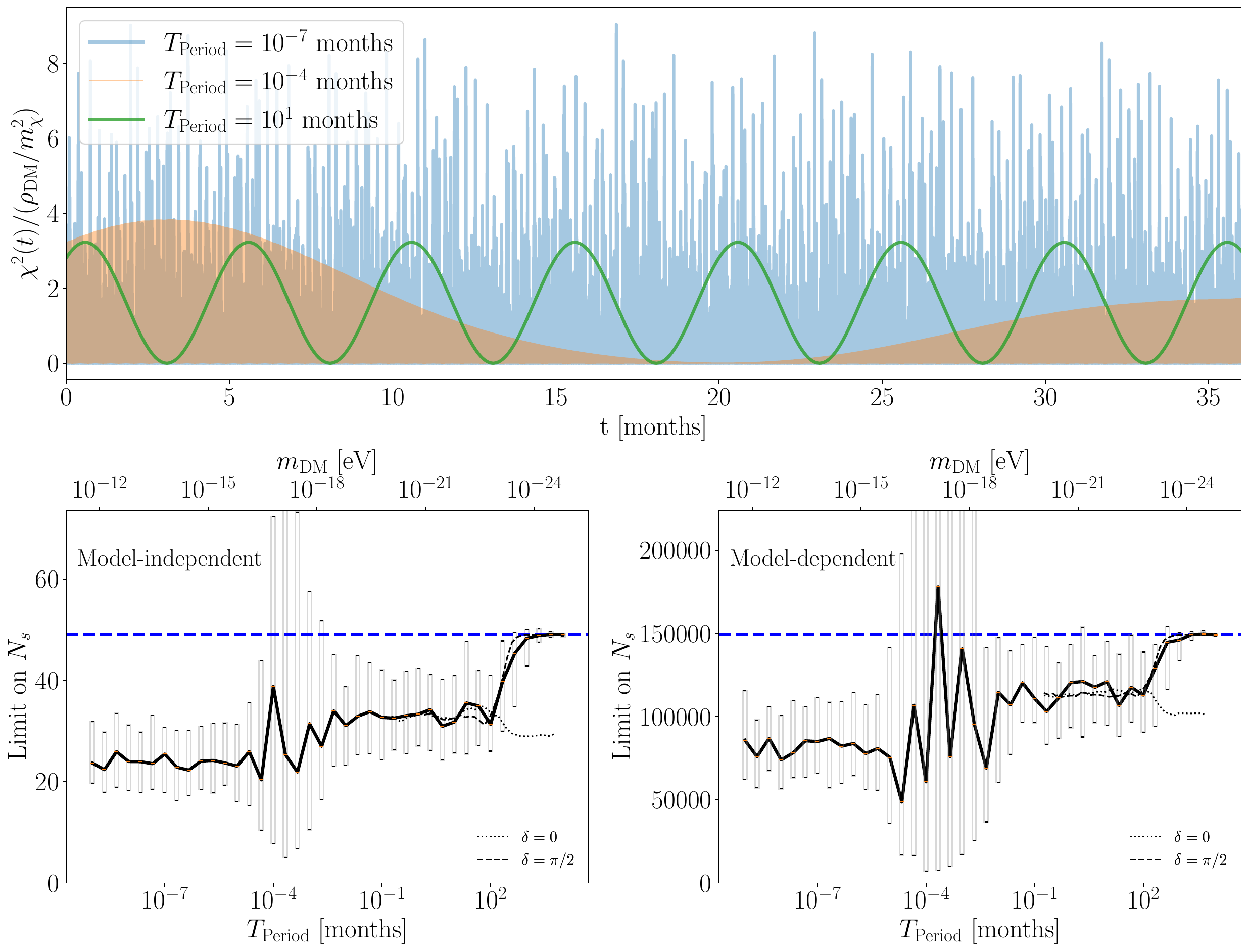}
  \caption{\textbf{Top}: An example of the time-dependent DM field normalized to the mean value, $[\chi(t)/(\sqrt{\rho_{\rm DM}}/m_\chi)]^2$. For the process shown in the upper diagram of \Cref{fig:diagrams}, the cross section is proportional to $\chi(t)^2$. For long oscillation periods, the time-dependence is well approximated as $\sin^2(2\pi t/T_{\rm Period} + \delta)$ over the duration of the experiment, in this case $36$ months. For shorter oscillation periods, $T_{\rm Period}\times10^6<36$ months, the velocity dispersion in the background DM causes the amplitude to vary with time, while retaining the rapid $T_{\rm Period}$ oscillation. The orange and blue curves are rapidly oscillating with the shown $T_{\rm Period}$, which cannot be resolved in the graphic---the slower oscillation of the amplitude corresponds to the coherence time.  \textbf{Bottom left}:  Median limit at 95 \% CL on the number of signal events vs $T_{\rm Period}$ (alternatively, $m_{\rm DM} = 2\pi/T_{\rm Period}$). The blue dashed line is the limit on the number of signal events $N_s$ without taking time-dependence into account, and the solid black line is the limit we find if the time-dependence is taken into account. This result is somewhat model-independent, as it only uses the observed background in a single kinematic bin from Ref.~\cite{ATLAS:2021kxv} which had a predicted background rate of $223\pm 19$. \textbf{Bottom right}: Similar to the bottom left panel, but for the model searched for in Ref.~\cite{ATLAS:2021kxv}, which gives the kinematic distribution that we assume has an underlying time-dependence proportional to $\chi^2(t)$. Here, $N_s$ is the number of signal events summed across all bins of missing transverse momenta $\slash{p}_T$. The constraining power is dominated by a single kinematic bin, with a predicted background rate of $219\pm 9$. For both bottom panels, in the dashed (dotted) line we also show for illustration the result obtained for $\delta=\pi/2$ $(\delta=0)$, which has important implications on the resulting limit for long oscillation periods. Also for both bottom panels, the vertical bars denote the 25\% and 75\% quartile ranges of the resulting constraint; the size of these bars is determined by a combination of the finite statistics and the unknown DM parameters that we sample over.} 
\label{fig:TimeDependence_pTLimts}
\end{figure*}

ATLAS sets limits on several dark matter models by searching for excesses at large missing transverse momentum, $\missET$, in events with at most several high-$p_\textrm{T}$ jets and no leptons~\cite{ATLAS:2021kxv} that were collected over a span of $\approx36$ months, with a total integrated luminosity of $139$ fb$^{-1}$.  Their analysis assumes no intrinsic time variation in the signal. In this section, we reinterpret their search and apply it to a time-varying signal.  We find tighter limits than those placed on models in which the signal is uniform in time.  We do this for both their model-independent limits, which places a limit using a single inclusive $\missET$ bin, and their model-dependent limit, which uses the shape of the $\missET$ spectrum.  In both cases, we take motivation from \Cref{ref:model}, which introduces time dependence from the classical field value of background dark matter, but we investigate a broader range of periods for the oscillation.  
The improved limits are primarily due to concentration of the signal in time, allowing for higher signal-to-background ratios.

\subsection{Model-independent limits}

ATLAS set model-independent limits on the number of signal events, $N_s$, that are compatible with the expected backgrounds and observed events as a function of a minimum $\missET$ threshold.  This inclusive single-bin analysis makes no further assumptions about kinematic distributions of the signal which survive the selection, relying only on event counts.  We choose  the highest $\missET$ bin\footnote{Region IM12 from Table VIII in Ref~\cite{ATLAS:2021kxv}.} which requires $\missET > 1200 \gev$. In this inclusive bin, the expected background is 223$\pm$19 events, while 207 events are observed. 

We assume that the background is time-independent\footnote{Some systematic effects that could break the time-independence of the background are discussed in \Cref{sec:discussion}.}, and distributed evenly throughout the 36-month run.  We add time dependence to the hypothetical signal following the time-dependent probability that varies as the square of the background DM field
\begin{align}
P_s(t) \propto \chi(t)^2~.
\end{align}

In this section we consider the range $T_{\rm Period}=2\pi/m_\chi \in [10^{-9}, 10^{3}]$ months. The upper panel of \Cref{fig:TimeDependence_pTLimts} shows several examples of the time dependence for different choices of the period and a given $\{\alpha_j,\delta_j\}$. For long oscillation periods, and thus even longer coherence times, the time-dependence is well approximated as $\propto \sin^2(2\pi t/T_{\rm Period} + \delta)$ over the duration of the experiment, in this case $36$ months. For shorter oscillation periods, $T_{\rm Period}\times10^6<36$ months, the velocity dispersion in the background DM causes the amplitude to vary with time, while retaining the rapid $T_{\rm Period}$ oscillation. Thus, there is both a short-time-scale oscillation with $T_{\rm Period}$ as well as modulation in amplitude which varies over a timescale of $\sim 10^6 \times T_{\rm Period}$. 

Limits are calculated by performing an unbinned-in-time likelihood analysis for events in a single $\missET$ bin. The extended likelihood is written for the $n=207$ events observed at times $\{t_i\}$ in the highest $\missET$ bin as
\begin{equation}
\label{eq:likelihood}
    {\cal L}(N_s,r) = \left[\frac{\mu^{n}e^{-\mu}}{n!} \prod_i^n f(t_i)\right]\times e^{-(r-1)^2/2\sigma_{r}^2}~,
\end{equation} 
\noindent
where $f(t)$ is the probability density function (pdf) formed from the weighted combination of the pdfs for $N_s ~(N_b)$ signal (background) events \textit{i.e.}~$f(t_i) = [N_s f_s(t_i) + N_b f_b(t_i)]/(N_s+N_b)$ and $\mu=N_s+N_b$ is the predicted number of events in the highest $\missET$ bin. The Gaussian factor with nuisance parameter $r$ describes the uncertainty on the prediction for the number of background events, such that $N_b\equiv 223r$ with the relative uncertainty on the number of background events from the ATLAS analysis, $\sigma_r=19/223$. For a given sampling of $n$ events that are drawn from a uniform distribution in time, the likelihood is maximized at $r\approx 1$. The maximum-likelihood estimate of $N_s$, on the other hand, depends on the particular set of drawn events, $\{t_i\}$, and on the exact form of $f_s(t_i)$, which itself depends on the parameters $\{\alpha_j,\delta_j\}$ from \Cref{eq:timedep}. Because the drawn events are uniformly distributed while the signal PDF is time-dependent, the likelihood is typically maximized at $N_s=0$. However, this is not always the case: for example, if the drawn event times happen to coincide with times where the signal pdf $f_s(t_i)$ is large, the likelihood can be maximized at $N_s>0$ (this is essentially the look-elsewhere effect). As we will discuss further below, such configurations are not common, but they have important consequences for certain regions of parameter space, as they yield upper bounds on $N_s$ that are weaker than those obtained from realizations in which the signal PDF does not coincide with the drawn events.

The time-dependent limit is compared to the time-independent limit from the ATLAS analysis, which we reproduce in our analysis by removing time dependence to leave a single-bin counting experiment
${\cal L}(N_s,r) = \mu^ne^{-\mu}/n!\times e^{-(r-1)^2/2\sigma_r^2}$, (\textit{i.e.} taking $f(t_i)\rightarrow1$) in \Cref{eq:likelihood}. Here, the likelihood is always maximized at $N_s=0$ because they observed a deficit with respect to the background. The fit is only determined by the total number of observed and expected background events with no fine-grained time structure that can point to the presence of a signal, as in the time-dependent analysis in the previous paragraph.

For each $T_{\rm Period}$, we sample over the parameters $\alpha_j$ and $\delta_j$ that determine the time dependence in \Cref{eq:timedep}, drawing a new set of $\{t_i\}$ for each sample. We then take the median limit obtained over 100 such realizations. Depending on $T_{\rm Period}$, the results can be very sensitive to different $\{\alpha_i,\delta_i\}$. This is especially for $T_{\rm Period}\approx36 ~{\rm months}/10^6$, where there is only one coherence period observed over the duration of the experiment as approximated by the orange curve in the top panel of \Cref{fig:TimeDependence_pTLimts} (note that this curve is rapidly oscillating and the slower oscillation is the drift of the amplitude). 

The model-independent limits on $N_s$  as a function of $T_{\rm Period}$ are shown in the lower left panel of \Cref{fig:TimeDependence_pTLimts}. The dashed blue line is the limit resulting from the analysis that does not include time information, and the solid black line shows the result obtained using time information. For very large oscillation periods, the result merges with the time-independent result, as it is effectively time-independent. The chosen phase can have important implications.  In particular, for $\delta \lsim 2\pi T_{\rm exp}/T_{\rm period}$ the time-dependence is approximately linear, which can be strongly constrained.  We show this case by singling out $\delta =0$ in the plots, as well as showing $\delta=\pi/2$ explicitly.

For $10^{-4}\lesssim T_{\rm Period}/{\rm month}\lesssim 36$, the time-dependence approaches the simple form of $\sin^2(t)$, and the resulting improvement on $N_s$ is a factor of about 1.6 over the time-independent analysis. For smaller oscillation periods, the coherence time plays an important role and results in a richer time-dependence. In particular, for $T_{\rm Period} \lesssim 10^{-4}$ months, the power distributed 
in the so-called ``slow mode''~\cite{Hui:2021tkt,Kim:2024xcr,Dror:2025nvg}, associated with the kinetic energy of the DM and 
characterized by a timescale $T_{\rm cohere.}$, can be resolved by the LHC over 
the observation time of 36 months. The overall result for such short periods is that the limit on $N_s$ is about a factor of two stronger than the time-independent analysis. However, it is worth noting that the level of improvement is strongly influenced by the uncertainty on the background according to $\sigma_r$ from \Cref{eq:likelihood}; we have tested the limit of a completely unknown background, $\sigma_r\rightarrow \infty$, and we find a larger relative improvement over the time-independent analysis of $\approx 4 ~(7)$ for $T_{\rm Period}\approx 1 ~(10^{-8})$ months because the time information allows for a more direct probe of the background distribution. For smaller oscillation periods below $10^{-9}$ months, it is not expected that the results will change significantly. However, for time-dependent signals with oscillation periods below the bunch spacing of the LHC, ${\cal O}(10 ~{\rm ns})$, the signal will appear to be time-independent, and the resulting limit on $N_s$ will merge back to the time-independent result.

We also show the interquartile range (25th to 75th percentile) over the 100 realizations as vertical bars for each $T_{\rm Period}$. The size of the range is driven by a combination of the finite number of observed events and the variance in the sampled DM model parameters $\alpha_j$ and $\delta_j$. The latter effect is most pronounced when the timescale of the modulation is comparable to the experimental exposure, \textit{i.e.}\ for $T_{\rm Period} \approx 36$~months (note the spread between $\delta = 0$ and $\delta = \pi/2$), or when $T_{\rm cohere.} = 10^6 \times T_{\rm Period} \approx 36$~months, corresponding to $T_{\rm Period} \approx 10^{-4}$~months. In this regime, the signal shape varies substantially with the parameters $\alpha_j$ and $\delta_j$, and certain draws of $t_i$ can mimic a signal-like distribution. This can cause the likelihood to be maximized at $N_s > 0$, as discussed previously, resulting in a weaker constraint than the time-independent analysis for some realizations which are only sensitive to the overall rate. In an experimental setting, such an outcome would suggest possible evidence for a signal, which could be confirmed by collecting additional events over a longer time. Moreover, in this region the more rapid oscillation set by the DM mass, with period $T_{\rm Period} = 2\pi/m_{\rm DM}$, would cluster the signal events in time and could ultimately be resolved with sufficient statistics.

\subsection{Model-dependent limits}

ATLAS also set limits on specific models which produce events with large $\missET$, such as those in which a dark vector boson $Z_A$ with a mass of 2 TeV decays into weakly-interacting massive particles of mass 1 GeV.  As the model provides a specific prediction\footnote{Referred to as DMA in Ref~\cite{ATLAS:2021kxv}; exclusive bin contents can be found in Table VII.} for the signal distribution in $\missET$, the entire observed $\missET$ spectrum can be used to set more powerful limits than the model-independent results above. Potential time dependence can be incorporated by extending the interaction studied in Ref.~\cite{ATLAS:2021kxv} to include a coupling to a time-varying SM gauge-singlet scalar field by inserting a factor of $\chi/\Lambda$ into the Lagrangian, which would have no effect on kinematic distributions if $\chi$ is taken to be the background field value.

Assuming a single monolithic background with a single uncertainty in each $\missET$ bin, we extend each exclusive $\missET$ bin by adding a time dimension. As above, we assume the background is time-independent and the signal is time-varying. 
Limits are calculated by extending the likelihood in \Cref{eq:likelihood} from the previous section across the multiple $\missET$ bins. There is one nuisance parameter for each bin, each of which is varied in an uncorrelated way.

The model-dependent limits are shown in the bottom right panel of \Cref{fig:TimeDependence_pTLimts}.  The improvement between time-independent and time-dependent is qualitatively similar to the model-independent results, though the constraint on the number of events is much larger due to the integration over all $\missET$ bins. We note that the improvement is driven by the last  $\missET$ bin, which has the largest expected signal-to-background ratio and an estimated number of background events $218\pm9$.

\section{Time-dependent resonances}

We now explore the power of time dependence to enhance searches for anomalous resonances which appear as localized signals in invariant mass over a smooth background.  In this section, for simplicity, we will only consider oscillation periods that result in coherence times which are much longer than the duration of the experiment. In this case, the DM is well-approximated by a sine wave of a single frequency that has a constant amplitude. For shorter periods and coherence times, the power of time dependence would likely get stronger, as in the previous section.

First, we consider a case when the form of the time dependence of the signal oscillation, including the period and phase, is assumed to be known. Then we consider a case when the parameters of the signal oscillation are learned from the data. Finally, we consider a third approach which utilizes sidebands in time where the signal is minimal to learn the nature of the background within the mass window of the resonant signal.

We assume a model of the background which allows for generation of Monte Carlo samples; the techniques shown here can also be applied to resonance searches in which the background is learned from the data, such as invariant mass sideband extrapolation.

For the first two considered cases of known and learned signal oscillation -- \Cref{sec:knownosc} and \Cref{sec:learnedosc} respectively -- we generate 1M QCD dijet background events and 100k signal events from the Lagrangian in \Cref{eq:lagrangian} for a scalar $\phi$ with $m_\phi=750$ GeV, decaying to jets.  For the third case where we study sidebands in time, \Cref{sec:time-sidebands}, we generate 100k signal events with $m_\phi=250$ GeV, again decaying to jets.

Collisions and decays are simulated with {\sc Madgraph5} v3.5.7~\cite{madgraph} using {\sc nnpdf} v2.3~\cite{NNPDF:2021uiq} for the parton distribution, and renormalization and factorization scales set to $m_Z$.  {\sc Pythia} v8.306~\cite{pythia} is used for fragmentation and hadronization. Radiation of additional gluons is modeled by {\sc Pythia}. The detector response is simulated with {\sc Delphes} v3.5.0~\cite{delphes} using the standard CMS card, and {\sc root} version 5.34.25~\cite{ROOT}.   
Narrow-cone jets are clustered using the anti-$k_{\textrm{T}}$ algorithm~\cite{Cacciari:2008gp} with radius parameter $R = 0.4$ using \textsc{FastJet 3.1.2}~\cite{Cacciari:2011ma} and are required to have $p_\textrm{T}\geq20$ GeV and $|\eta|\leq2.5$.  Events are required to have at least two jets, and the invariant dijet mass is reconstructed from the two jets with the highest $p_\textrm{T}$.

\subsection{Known oscillation}
\label{sec:knownosc}
\begin{figure}
    \centering
    \includegraphics[width=0.9\linewidth]{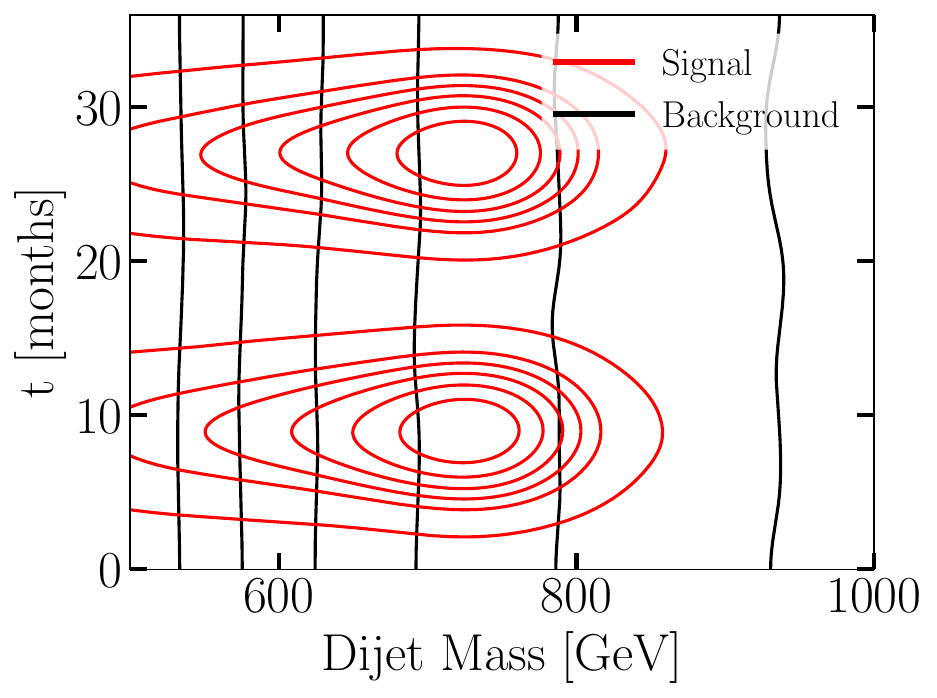}
     \includegraphics[width=0.83\linewidth]{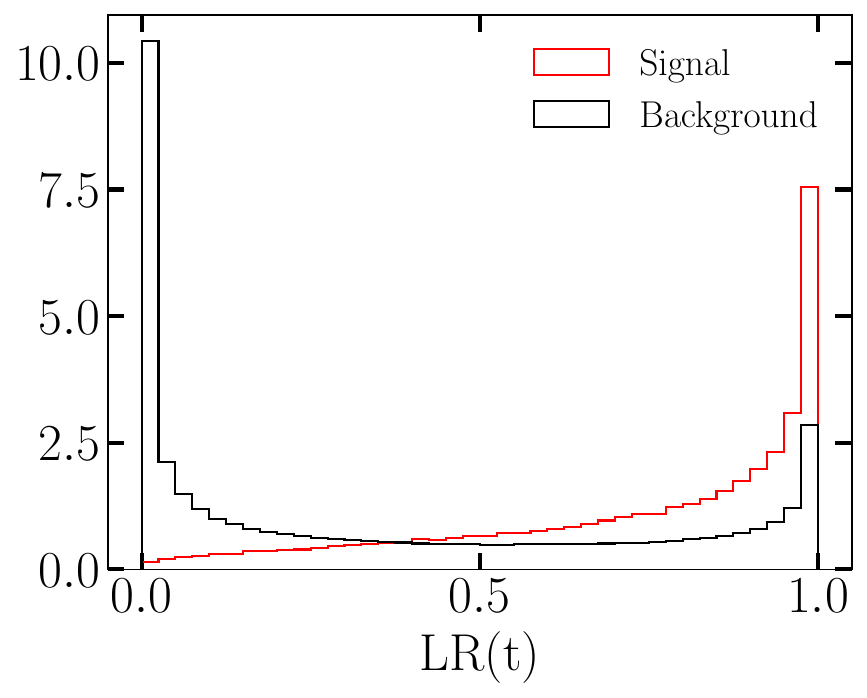}
    \caption{\textbf{Top}: Distribution of masses versus time for signal $\phi$ and background QCD dijets. The background is time-independent, wiggles are statistical. \textbf{Bottom}: LR distribution for signal and background, showing the expected separation.}
    \label{fig:res_time}
\end{figure}

The likelihood ratio (LR), or any monotonic function of it, is the optimal tool for simple hypothesis testing~\cite{Neyman:1933wgr} and is widely employed in practice (though improvements exist for the general case~\cite{Carzon:2025isu}).
Traditionally, the LR is expressed in terms of the kinematic properties of the signal and background. In a dijet resonance search, the LR is often factorized into pieces which depend on the dijet invariant mass, $m_{jj}$, and auxiliary components. Applying a selection on the other components can enhance the signal-to-background ratio, enabling a more powerful hypothesis test in the mass feature.
If the signal and background have different temporal dependence, then the auxiliary term of the LR contains a time-dependent factor.  If the functional form of the temporal oscillations is known, then the time component of the LR can be factorized as:
\begin{equation}
    {\rm LR}(t) = \frac{p_s(t)}{p_b}.
\end{equation}
This can calculated analytically using the same time-dependence assumed above---for a resonance, we assume a $p_s(t)\propto \sin^4(2\pi t/T_{\mathrm{period}}+\delta)$ time-dependence. In the upper panel of \Cref{fig:res_time}, we show contours of constant event rate in the space of dijet invariant mass and time for signal (red) with an oscillation period of 36 months and background (black) with a uniform distribution in time. In the lower panel, we evaluate the likelihood ratio for the events from each component, and plot their normalized distribution. 

One common approach to using the auxiliary component of the LR is to apply a selection, such as a minimum threshold, to suppress the background relative to the signal. However, this sacrifices information, as removed signal events cannot contribute and removed background events cannot be used to profile nuisance parameters in the background model. Instead, we weight events by the LR to maximize use of information~\cite{Gambhir:2025afb}. 

\begin{figure}
    \centering
    \includegraphics[width=0.95\linewidth]{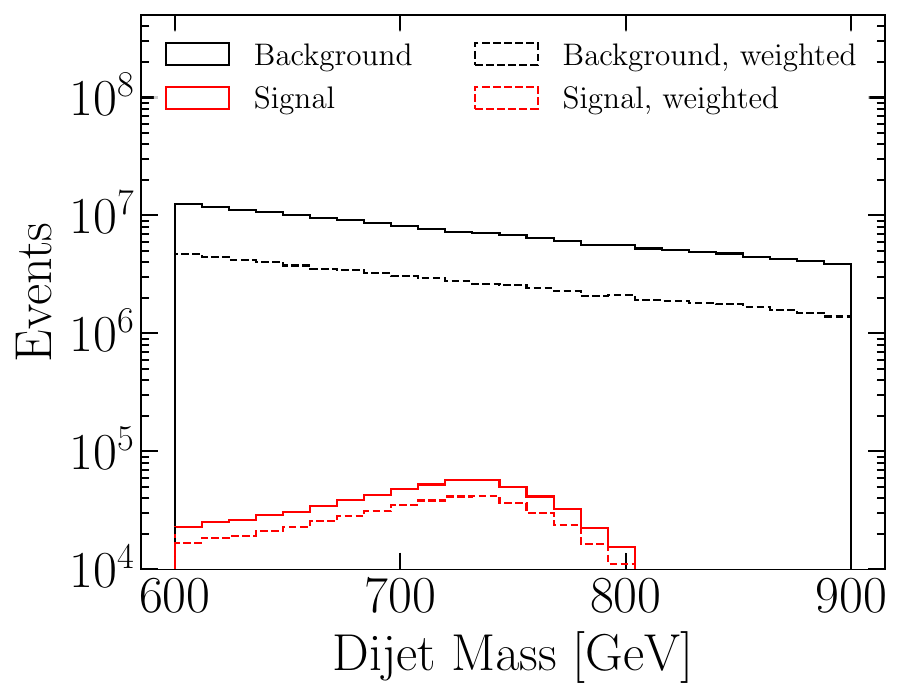}
    \includegraphics[width=0.945\linewidth]{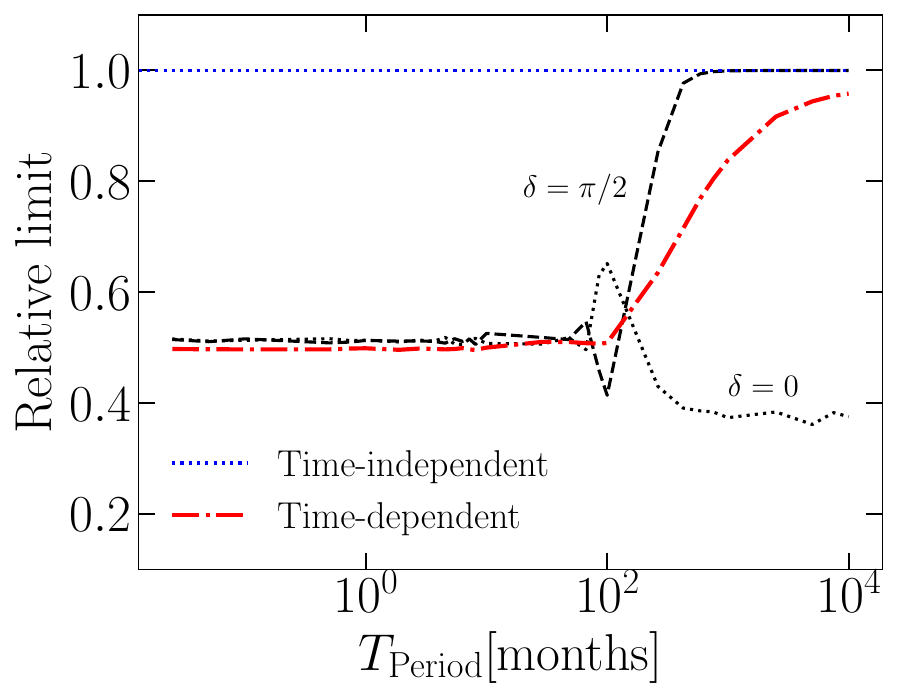}
    \caption{\textbf{Top}: Mass distribution of the signal (red) and background (black) before (solid) and after (dashed) weighting by the LR. \textbf{Bottom}: Expected limits versus $T_{\rm Period}$ with (red) and without (blue) using time-dependent information for an assumed $T_{\rm Period}$}.
    \label{fig:res_time_results}
\end{figure}

In this study, the dijet background is normalized to match the observed spectrum at CMS~\cite{CMS-PAS-EXO-23-004}, such that the total number of events is the same in the invariant dijet mass region $660 \leq m_\phi/{\rm GeV}\leq 900$. The signal scale factor $\mu=1$ is defined to give 1M events in the mass window.
Application of the time-based weights suppresses the background more than the signal; the upper panel of \Cref{fig:res_time_results} shows the distribution of events in bins of invariant dijet mass before and after weighting by the likelihood ratio for signal and background. This allows a more powerful extraction of the signal than if the time information is not used.  

Upper limits are calculated for the time-varying and time-independent scenario assuming a coherent 10\% background uncertainty before profiling on the sidebands. Limits are calculated with {\sc PyHF}~\cite{pyhf,pyhf_joss} under the assumption of the asymptotic regime~\cite{Cowan:2010js} and are presented in the lower panel of  Fig~\ref{fig:res_time_results}. For periods between $10^{-2}$ and 1 month, time-dependent limits are approximately 50\% stronger than those that do not use temporal information. 

\subsection{Learned oscillation}
\label{sec:learnedosc}

The above analysis requires knowing the time dependence in advance, or at a minimum scanning the parameters of the assumed oscillation. In this section, we demonstrate a method for discovering a time-dependent resonance without specifying a single target period or phase in advance. We assume a simple time-independent background; in a more realistic treatment beyond the scope of this initial exploration, one could consider time-dependent effects in the background which might mimic signal temporal dependence. We discuss this further in \Cref{sec:discussion}.

The CATHODE~\cite{Hallin:2021wme} method is a localized-anomaly-detection technique which learns a score function that summarizes the separation power of all auxiliary features (\textit{i.e.}~those other than the feature used to define sidebands). A model of the background in all auxiliary features is learned as a function of mass from events in the mass sidebands, and used to generate events that describe the background within the mass signal window. After generating background events, CATHODE trains a weakly-supervised classifier to distinguish data from the background model within the signal window using the auxiliary features. In this way, the classifier learns an LR-like time score which we will denote ${\rm LR}(t)$.

In this application, the signal region is defined by
$m_{jj}\in[550,850]~\mathrm{GeV}$.
The time information of the simulated events is used as the auxiliary input; we include the timestamp of the events (normalized to the experimental run time, $t_{\rm exp}$), as well as a set of Fourier features:
\begin{equation}
\begin{aligned}
\mathbf{x}_t
= \Bigg[
\frac{t}{t_{\rm exp}},\;
\Big\{
\sin\!\left(2\pi h\,\frac{t}{T_j}\right),\,
\cos\!\left(2\pi h\,\frac{t}{T_j}\right)
\Big\}_{\substack{j=1,\ldots,100\\ h=1,\ldots,4}}
\Bigg],
\end{aligned}
\end{equation}
with
\begin{equation}
\nonumber
t_{\rm exp}=36~\mathrm{months},\quad
T_j\in[0.1,10^4]~\mathrm{months}.\quad
\end{equation}
Here, $h\leq4$ is motivated by the fact that the processes we are interested in have rates that depend on (up to) the fourth power of the classical field and $\sin^4(x)=[3-4\cos(2x)+\cos(4x)]/8$.  Being more general would require going to higher values of $h$. Here, $T_j$ are uniformly distributed in log-space over the region of interest.

\begin{figure}
    \centering
    \includegraphics[width=0.99\linewidth]{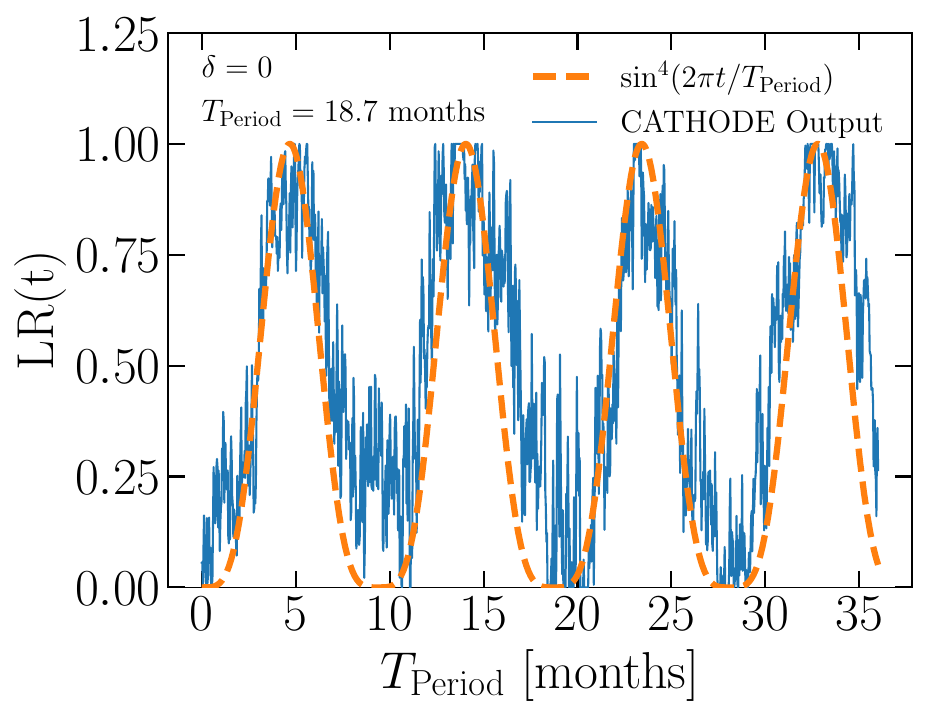}
        \includegraphics[width=0.95\linewidth]{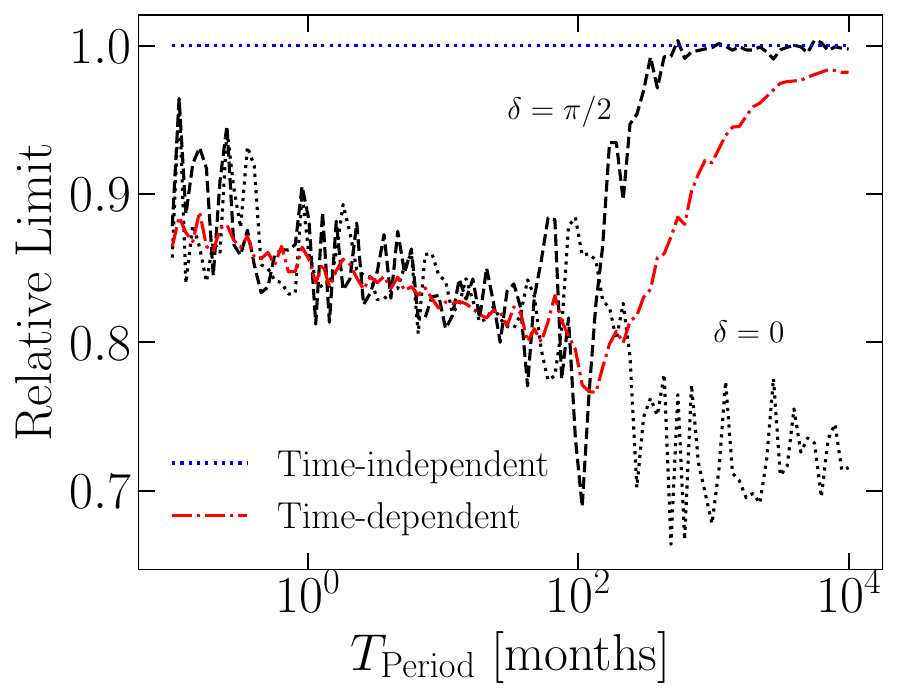}
        
    \caption{\textbf{Top:} CATHODE learned time score (solid blue) compared to the true temporal modulation (dashed orange). CATHODE is trained without injecting the true period or phase and without imposing the exact target waveform; instead, it learns an LR-like discriminator using a fixed Fourier feature bank. \textbf{Bottom:} The limit on the number of signal events relative to the time-independent analysis.}
    \label{fig:cathode}
\end{figure}

The Fourier feature representation allows the classifier to learn periodic structure without pre-specifying the true period or phase (within the chosen basis range). The upper panel of \Cref{fig:cathode} compares the learned time score to the true time dependence for $T_{\rm Period}=18.7$ months at a fixed phase $\delta=0$, showing that CATHODE can capture discriminating timing information. The CATHODE classifier is then used in place of the analytic LR from \Cref{sec:knownosc} to define event weights and enhance signal relative to background.

We are now in a position to set expected limits. In the lower panel of \Cref{fig:cathode}, we show the corresponding expected limit relative to the time-independent limit using the learned event weighting. As expected, the bound is not quite as strong as in the known-oscillation case, but remains stronger than the time-independent bound. CATHODE struggles for rapid oscillations (small $T_{\rm Period}$), due to the difficulty for neural networks to learn high-frequency patterns, while for very long periods the modulation is weak within the finite exposure window and the bound approaches the time-independent result. Again we see the behavior mentioned earlier that small phases (in the plot we show $\delta=0$) can be more strongly constrained, but this would require a lucky coincidence in the actual experiment.

\subsection{Sidebands in time}
\label{sec:time-sidebands}

\begin{figure}
    \centering
    \includegraphics[width=0.95\linewidth]{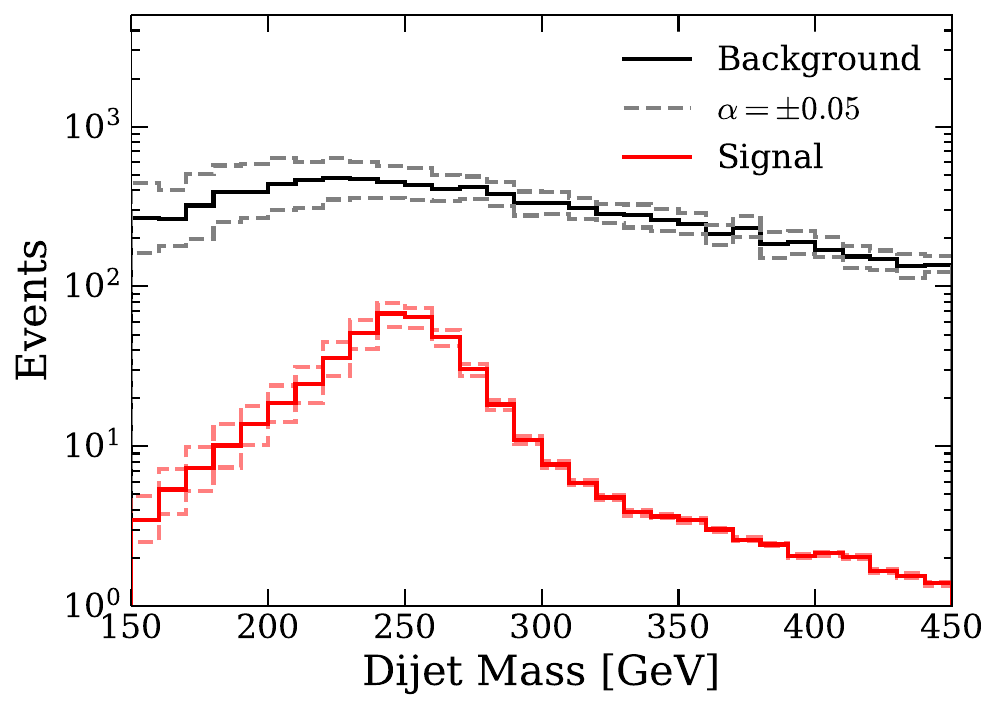}
    
    \caption{A  model of a localized signal in an invariant mass where the background has a non-trivial feature due to an efficiency turn-on. }
    \label{fig:timebandsv}
\end{figure}

 Resonance searches typically model the background in the signal region from background-dominated data samples to avoid reliance on simulation, often using sidebands in mass away from the resonance.  Interpolating into the signal region usually assumes a simple mass dependence for the background, but in many cases the actual dependence is complex, potentially including peaks due to kinematic thresholds or efficiency turn-ons.
 As an illustrative example which we will discuss further, Figure~\ref{fig:timebandsv}, shows an example of binned signal and background distributions, where the  background is not simply falling, but exhibits a broad peak near the resonance mass.
 In this scenario, modeling the background under the resonance peak using only sideband information is very challenging, as the sidebands are consistent with different background shapes in the signal region, as described by the uncertainty envelope. 

However, if the signal has a time-dependent behavior, then time intervals with low signal-to-background ratios can be as useful as those with high ratios, by providing an opportunity to learn the mass dependence of the background inside the signal region. The time dependence of the signal therefore allows for {\it sidebands in time} as well as mass, which can constrain the uncertainty on the data-driven background model. Sidebands in mass require selection of the signal region in advance, or scanning several regions; sidebands in time similarly require advance knowledge of the time dependence, or scanning.

To construct the distributions shown in \Cref{fig:timebandsv}, we apply a turn-on efficiency curve to QCD dijet background events and signal events with a scalar $\phi$ with $m_\phi=250$ GeV, decaying into jets. The turn on depends on the leading jet $p_\textrm{T}$, as

\[ P(p_\textrm{T}) = [1 + e^{-(p_\textrm{T}-\mu_t(1+\alpha))/\sigma_t}]^{-1} \]

\noindent where $\mu_t=120$ GeV represents the turn-on threshold which also depends on a nuisance parameter $\alpha$ that represents uncertainty in the turn-on; uncertainty on $\alpha$ translates into uncertainty on the location of the background peak. 
We treat $\alpha$ as a nuisance parameter that has a Gaussian prior centered at 0 with width $\sigma_\alpha$, which will serve as our systematic uncertainty.
The width of the turn on is described by  $\sigma_t=10$ GeV.

\begin{figure}[t]
  \centering
  \includegraphics[width=0.95\linewidth]{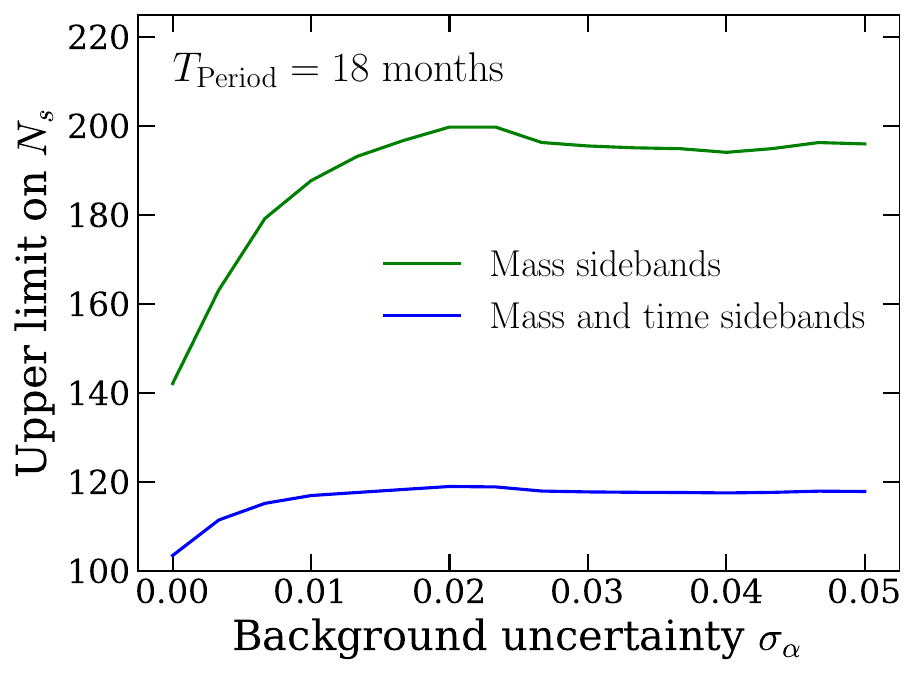}
  \includegraphics[width=0.95\linewidth]{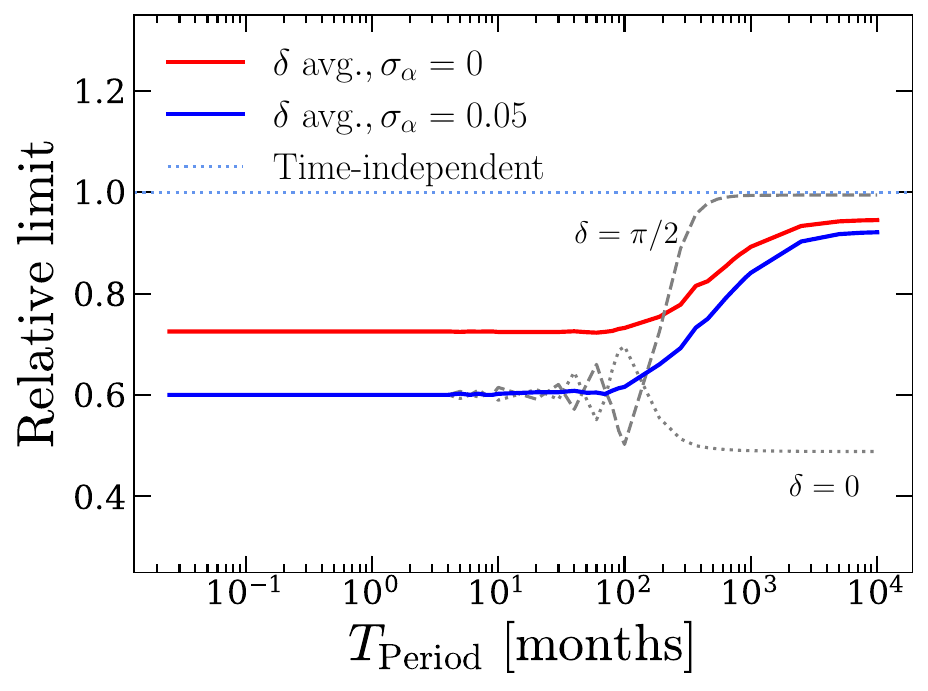}
  \caption{ \textbf{Top:} upper limit at 95\% CL on the number of signal events $N_s$ on top of the peaking background described in Fig.~\ref{fig:timebandsv}, for statistical analysis that uses only mass sidebands compared to analysis that uses mass and time sidebands to constrain the background model uncertainty. Shown as a function of $\sigma_\alpha$, the uncertainty on $\alpha$, which controls the location of the peak in the background model.  \textbf{Bottom:} The limit on the number of signal events for the analysis that uses time and mass sidebands relative to one that only uses mass sidebands as a function of $T_{\rm Period}$ and averaged over the random phase $\delta$. The red curve shows the improvement for the case with no systematic uncertainty, $\sigma_\alpha=0$, and the blue curve shows the case where $\sigma_\alpha=0.05$. For the latter case, we also show the improvement for a fixed $\delta=0$ ($\delta=\pi/2$) in black dotted (dashed) lines. }
  \label{fig:time-sidebands-scan-v}
\end{figure}

Upper limits on the number of signal events $N_s$ are set using {\sc PyHF} from a binned histogram in mass only, where the background  and signal models include the shape uncertainty described above.  When the  uncertainty is removed ($\sigma_\alpha= 0$), the 95\% CL upper limit on $N_s$ is approximately 150 events. When the uncertainty is increased, the statistical power fades significantly, to $N_s<200$ events with the trigger threshold uncertainty  of $\sigma_\alpha=0.05$; see the green curve in the upper panel of Fig.~\ref{fig:time-sidebands-scan-v} which assumes $T_{\rm Period}=18$ months and $\delta=0$. This highlights the importance of constraining the background model uncertainty in the signal region.

If the histograms are extended such that each mass bin is sub-divided in time, effectively adding sidebands in time, the statistical power improves. The case where $\sigma_\alpha=0$ already sees improvement due to the increase of signal-to-background ratio for periods of high signal amplitudes, from $N_s<150$ to $N_s<100$.  But when the background uncertainty is increased, the sensitivity remains robust, only weakening slightly to $N_s<120$ events; see the blue curve in the upper panel of Fig.~\ref{fig:time-sidebands-scan-v}.  This demonstrates the power of the low-signal-to-background regions to constrain uncertainty in the background model.

A scan of period and phases shows a dependence similar to other cases studied here; see the lower panel of Fig.~\ref{fig:time-sidebands-scan-v}. Here, the improvement gained by using time information is larger for $\sigma_\alpha=0.05$ than $\sigma_\alpha=0$ because the sidebands in time allow for the background in the resonance region to be directly constrained.

\section{Discussion of backgrounds}\label{sec:discussion}

Time-dependent analyses have the power to greatly increase the sensitivity to new physics that has time-dependent effects. So far, for simplicity, we have ignored the possibility that the backgrounds have some time dependence.  In a full analysis, with access to actual collider data, these time-dependent systematic effects must be taken into account.  We now consider some likely candidates for time-dependent backgrounds and discuss ways to mitigate them.  In some cases the time dependence of the systematic is not oscillatory in nature and should be easily separated from signal.  While in other cases the systematic is oscillatory and will present an irreducible background and may limit which signal frequencies the experiment is sensitive to.

 Systematic effects stemming from beam dynamics will result in a non-uniform time-dependence. Trivially, the bunch spacing of $\approx 25~{\rm ns}$ produce events in the detector on the same frequency. Over the course of operation, the instantaneous luminosity degrades over time on a timescale of ${\cal O}(10)$ hours~\cite{Kostoglou:2019yel,Lamont:1967453}, so the collision rate smoothly falls on the same timescale. A variety of dust particles contribute to beam losses throughout the LHC with characteristic time profiles~\cite{Lindstrom:2020hks}; some types of dust particles are localized to a certain location on the ring and thus contribute losses with an oscillation period corresponding to the orbital time $\approx 90 ~\mu{\rm s}$. The precise position of the luminous region at the interaction point changes with time shortly after a proton fill, and on longer timescales on the scale of months~\cite{ATLAS:2016nnj}. Beam-beam interactions~\cite{Babaev:2023fim,Papotti:2014dpa}, ground motion of natural and human-made origins~\cite{Schaumann:2023gst}, changes in detector performance due to radiation~\cite{Gibson:2011nma,ATLAS:2021gld} can all impart time-dependence that may mimic time-dependent new physics signals, although in many cases this time dependence is not expected to be periodic over long durations.  
 
 While we have mentioned a few important systematic effects, this list is not meant to be exhaustive and a dedicated study must be undertaken to identify all sources of time-dependent background. In practice, this analysis could be done in Fourier space, where a signal would appear as a peak and would need to be confirmed whether or not it is background.  For the case of a signal (e.g.~monojet) which varies as $\chi(t)^2$ this analysis can be done in an unbinned way using the Rayleigh periodogram \cite{Losada:2023zap,2016ApJ...822...14B}. 
 Note that for the DM origin of BSM time-independence, the signal frequency peak would have a width associated with the local DM velocity which could be useful for differentiating the signal from the background.

\section{Conclusions}

The probability of observing signals from certain models of new physics may vary with time. The most prominent example is that of an ultralight DM component that interacts with SM fields. This class of interaction is realized in a variety of models and is typically explored at a number of low-energy experiments. However, if the interaction is restricted such that the SM only interacts with the DM in the presence of an additional TeV-scale particle, then experiments at the energy frontier are more appropriate. Current analyses at the LHC assume that new signals do not exhibit any intrinsic time-dependence, such that the hypothetical signal rate is uniformly distributed in time up to systematic effects from the machine.

In this article, we explore the statistical power that can be gained if time information is used for interactions that exhibit non-trivial time-dependence. Up to systematic effects, the SM background will be time-invariant. This provides a powerful discriminator to separate signal from background and results in tighter constraints on the signal rate from new physics. 

We first consider a completed ATLAS analysis which searched for missing energy from stable final states that are produced at the LHC, and suppose that the hypothetical signal exhibits an oscillatory time-dependence corresponding to the DM mass with a period $T_{\rm Period}=2\pi/m_{\rm DM}$. For very long $T_{\rm Period}$ with respect to the experimental exposure, the timing information does not provide any statistical power as the signal is effectively time-independent; for shorter $T_{\rm Period}$ on the other hand, the limit on the number of signal events can be improved by a factor of $\approx 2$ if the timing information is used. Furthermore, because time information provides a handle to disentangle the time-independent SM background from a time-dependent signal, the utility of incorporating time information is sensitive to the uncertainty on the SM background prediction. In the limiting case of a completely unconstrained background, we find that this improvement factor is $\approx$ 7. In this way, other LHC analyses that have a less well-understood background can expect a large improvement in sensitivity when looking for a time-dependent signal if time information is used.

We further explore the power of time information for resonance signals. We first consider the case where the time-dependence is known and use the time component of the known likelihood ratio to reweight events. Signal events receive a larger weight than background, and resulting limits are produced and found to be stronger than the case where time information is not used. We extend this analysis by considering the case where the oscillation period is unknown, and use the CATHODE classifier to learn the time-dependence using motivated Fourier features which do not assume the oscillation period. Finally, we extend the usual ``sidebands" method of extracting the background in an invariant mass distribution to include the time direction.  This allows for direct access to the background rate in the invariant-mass signal region. 

Several points deserve further comment. In this work, we have considered the case in which the DM component enters as a classical background field. However, the same particles could also be produced directly at the LHC, yielding time-independent events with distinct kinematic distributions and signatures that may offer promising search channels. A key distinction is that a signature relying solely on collider-produced DM cannot confirm that the particle constitutes the cosmological dark matter. The optimal search channel will depend on the BSM model under consideration, including the relevant backgrounds and overall signal rate; for ambient DM, the latter is likely enhanced (relative to the collider produced DM) due to the large occupation number of the ultralight field. More broadly, within an effective field theory framework, any interaction can acquire time dependence through the insertion of a gauge-singlet ultralight DM field in the Lagrangian. For such an interaction, an important theoretical question is whether the ultralight DM mass is protected by a symmetry or whether keeping it ultralight requires fine-tuning.

For the novel search strategy we propose, it is essential to account for time-dependent collision rates arising from the many systematic effects. While we outline several such effects, to our knowledge no dedicated study has been performed. For the signals considered in this work, a natural approach is to perform a Fourier analysis of the time-dependent event rates, treating both systematic variations and potential signals as distinct frequency modes. If this challenge can be overcome, time-dependent signals at the LHC present a promising opportunity to fully harness the power of the LHC to discover new physics or even make a direct detection of DM.

\section{Acknowledgments} DW is funded by the DOE Office of Science. The work of MF was supported by NSF Grant PHY-2210283 and was also supported by NSF Graduate Research Fellowship Award No. DGE-1839285. MF and PF are also supported by Fermilab which is administered by Fermi Forward Discovery Group, LLC under Contract No. 89243024CSC000002 with
the U.S. Department of Energy, Office of Science, Office of High Energy Physics.
The authors thank Asher Berlin, Henry Frisch and David Miller for useful discussions.

\appendix

\bibliography{timedep}

\end{document}